\documentclass[twocolumn,10pt]{IEEEtran}
\usepackage{xcolor}
\usepackage{amsfonts}
\usepackage{amssymb}
\usepackage{bm}
\usepackage{bbm}
\usepackage{mathrsfs}
\usepackage{relsize}
\usepackage{mathtools}
\usepackage{makecell}
\usepackage{algorithm}
\usepackage{multirow}
\usepackage{algpseudocode}
\usepackage[colorlinks,urlcolor=blue,linkcolor=black,citecolor=blue,breaklinks=true]{hyperref}
\usepackage{url}
\usepackage{subfigure}

\usepackage{cite}
\ifCLASSINFOpdf
  \usepackage[pdftex]{graphicx}
\else
\fi
\usepackage{amsmath}

\begin{document}

\title{Distributed Coordination for Heterogeneous Non-Terrestrial Networks}

\author{Jikang~Deng,~\IEEEmembership{Student Member,~IEEE,} Hui~Zhou,~\IEEEmembership{Member,~IEEE,} and Mohamed-Slim~Alouini,~\IEEEmembership{Fellow,~IEEE}
		
		\thanks{Jikang~Deng, and Mohamed-Slim~Alouini are with Computer, Electrical and Mathematical
Science and Engineering Division, King Abdullah University of Science and Technology (KAUST), Thuwal, 23955-6900, Kingdom of Saudi
Arabia (email: \href{mailto:jikang.deng@kaust.edu.sa}{jikang.deng@kaust.edu.sa};
\href{mailto:slim.alouini@kaust.edu.sa}{slim.alouini@kaust.edu.sa})}
		\thanks{Hui Zhou is with Centre for Future Transport and Cities, Coventry University, U.K. (email:\href{mailto:hui.zhou@coventry.ac.uk}{hui.zhou@coventry.ac.uk}). This work was done while he was working at KAUST.}
		
	}


\maketitle

\begin{abstract}
To achieve global coverage and ubiquitous connectivity, the non-terrestrial network (NTN) has been regarded as a key enabler in the sixth generation (6G) network, which includes uncrewed aerial vehicles (UAVs), high-altitude platforms (HAPs), and satellites. 
Since the unique characteristics of various NTN platforms strongly affect their implementation and lead to a highly dynamic and heterogeneous NTN scenario, achieving distributed coordination remains an important research direction. However, the explicit and systematic analysis of the individual layers' challenges and corresponding distributed coordination solutions in heterogeneous NTNs has not been proposed yet. Therefore, in this paper, we summarize the unique characteristics of each NTN platform, identify communication challenges within individual layers, and propose potential delay-tolerant or delay-sensitive coordinated solutions accordingly. We further analyse the feasibility of leveraging multi-agent deep reinforcement learning (MADRL) algorithms to achieve the proposed coordinated solutions. Finally, we present a case study of the joint scheduling and trajectory optimization problem in heterogeneous NTN, where a two-timescale multi-agent deep deterministic policy gradient (TTS-MADDPG) algorithm is developed to validate the effectiveness of distributed coordination.



\end{abstract}


%
\maketitle

\section{Introduction}
With the improved coverage and resilience against terrestrial infrastructure failures, the non-terrestrial network (NTN) has been identified as one of the emerging usage scenarios in International Mobile Telecommunications 2030 (IMT-2030) for the Six Generation (6G) network, which has the potential to provide connections in post-disaster scenarios and remote areas \cite{Xiao2024}. Currently, the research in the NTN community focuses on three platforms, including the uncrewed aerial vehicle (UAV), high-altitude platform (HAP), and satellites. It is noted that the unique characteristics of each platform (e.g., power supply and loading capability) play a critical role in the design and implementation of the NTN (e.g., transmission power and number of antennas) \cite{Toka2024}. Therefore, the coordination of heterogeneous NTNs remains an important challenge to solve.

Different from traditional terrestrial networks with fixed infrastructure, the NTN is characterized by dynamic network topology, which imposes a heavy burden on the centralized coordination among NTNs within stringent latency. In \cite{Tuzi2023}, the authors exploited the distributed homogeneous satellites to form a sparse phased array for direct-to-cell connectivity. Distributed learning has been regarded as a promising solution to enhance coordination among NTNs by capturing their dynamic topology, and several existing works mainly focus on the general distributed deep reinforcement learning (DRL) framework. In \cite{Xu2024}, the authors presented and analyzed buffer-aided NTN based on the decentralized DRL algorithm. In \cite{Cao2024}, the authors proposed generalized multi-tier DRL architectures to enhance the cooperation among space-tier, air-tier, and ground-tier stations. However, a comprehensive study of unique challenges and coordinated solutions in individual layers with distributed learning frameworks has never been carried out. 
\begin{table*}[!htb]
    \caption{Typical Characteristics of Non-terrestrial Network platform}
    \begin{center}
        \begin{tabular}{|c|c|c|c|c|c|c|}
            \hline
            \textbf{Specification}  &Untethered UAV&Tethered UAV&HAP&LEO&MEO&GEO\\
            \hline			
            Loading Capability &2.7 kg&20 kg& 140 kg&-&-&-\\
            \hline
            Power Supply &0.3 kWh&30 kWh&20 kWh&-&-&-\\
            \hline
            Latency  &1 ms&1 ms&1 ms&30-50 ms&150 ms&600 ms\\
            \hline
            Endurance &30 mins&24 hours& 2 months&5-7 years&10-15 years& 15-20 years\\
            \hline
            Deployment Cost&Low&Low& Medium&High&High&High\\
            \hline
            Network Type  &1 Micro/Pico &1 Macro &7 Multiple Macro &61 Multiple Macro &61 Multiple Macro &61 Multiple Macro \\
            \hline
            Payload Option & $\bullet$RU+DU+CU&\makecell[l]{$\bullet$ RU\\$\bullet$RU+DU\\$\bullet$RU+DU+CU}& \makecell[l]{$\bullet$RU+DU\\$\bullet$RU+DU+CU} &\makecell[l]{$\bullet$RU+DU\\$\bullet$RU+DU+CU}&\makecell[l]{$\bullet$RU+DU\\$\bullet$RU+DU+CU}&\makecell[l]{$\bullet$RU+DU\\$\bullet$RU+DU+CU}\\
            \hline
        \end{tabular}
    \label{aerial_BS_summarization}
    \end{center}
\end{table*}


Motivated by this, in this paper, we provide a concrete vision of coordination among heterogeneous NTNs with a focus on the corresponding distributed framework. The main contributions of this paper are as follows:
\begin{enumerate}
    \item We first summarize the unique characteristics of various NTN platforms, including UAV, HAP, and satellites, compared to the traditional terrestrial infrastructure, and analyze the corresponding impact on the design and implementation of NTNs in Section \ref{section:characteristic_NTN}.
    \item We provide a comprehensive analysis of challenges and coordinated solutions in heterogeneous NTNs, where the physical layer, MAC layer, network layer, and application layer are considered in Section \ref{section:challenges_solutions}.
    \item We further present emerging multi-agent deep reinforcement learning (MADRL) algorithm frameworks tailored for coordinated heterogeneous NTNs in Section \ref{section:distributed_learning}, which are divided into delay-tolerant and delay-sensitive solutions, respectively.
    \item We present a case study of joint optimization on user scheduling and trajectory design in a heterogeneous NTN network composed of tethered-UAV (T-UAV) and untethered-UAV (U-UAV), where the two-timescale multi-agent deep deterministic policy gradient (TTS-MADDPG) algorithm is utilized to validate the effectiveness of our proposed distributed coordination method. Finally, we conclude the paper in Section \ref{section:conclusion}.
\end{enumerate}
 
\section{Unique Characteristics of Non-terrestrial Network Platforms} \label{section:characteristic_NTN}
In this section, we present and analyze the characteristics of different aerial platforms, which include the UAV, HAP, and satellite. More importantly, we analyze how their unique characteristics impact the design and implementation of the cellular network as summarized in Table~\ref{aerial_BS_summarization}. 

\begin{figure}[h!]
  \centering
  \includegraphics[width = 8.8cm]{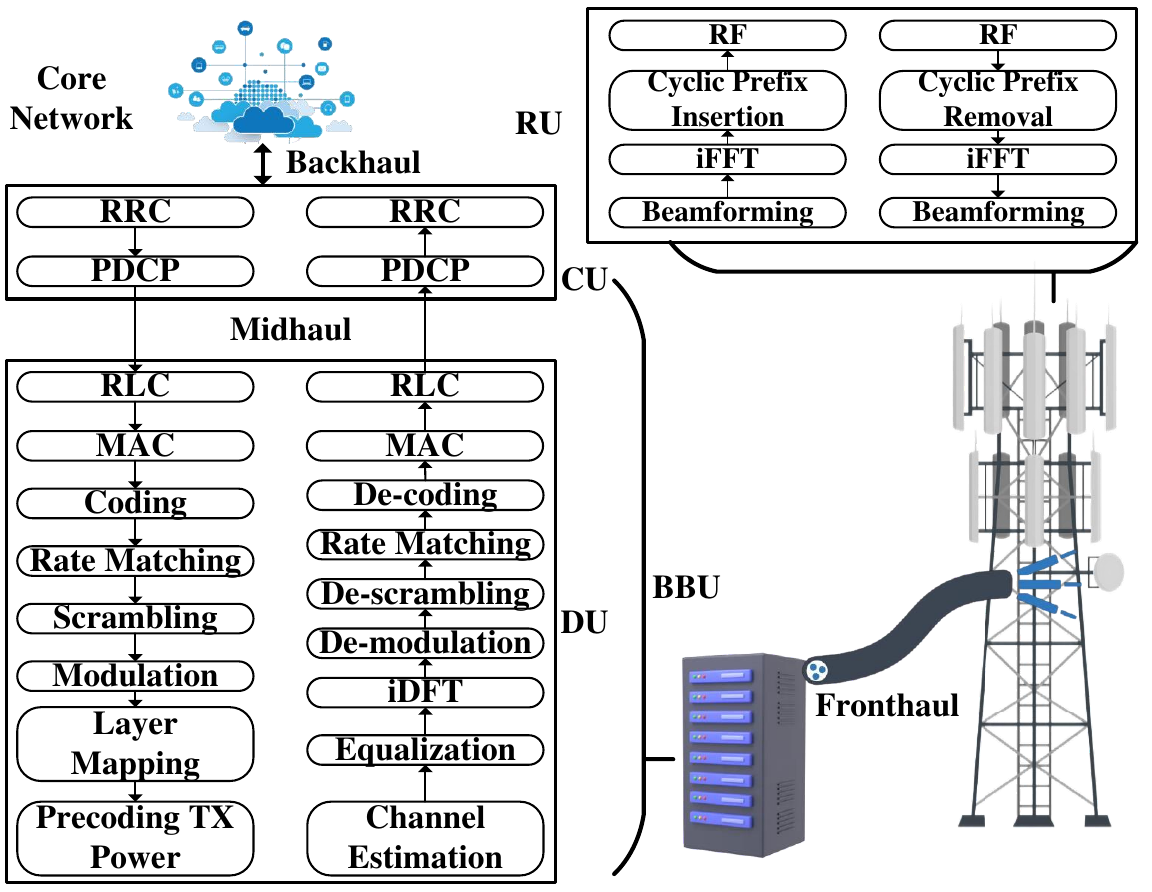}
  \caption{A typical structure of terrestrial cellular network}
  \label{aerial_BS_system}
\end{figure}

Fig. \ref{aerial_BS_system} shows the typical terrestrial 5G network consisting of a radio unit (RU), a distributed unit (DU), a centralized unit (CU), and the core network, where open radio access network (O-RAN) CatA is adopted for the fronthaul split \cite{Larsen2019}. To achieve the tradeoff between fronthaul bandwidth and coordination, the beamforming procedure, inverse fast Fourier transform (iFFT), and cyclic prefix processing are integrated within the RU, where the user data symbols and precoding weights are transmitted over the fronthaul. In terms of the size, weight, and power (SWaP),  it is noted that the typical weights of macro massive MIMO RU, and baseband unit (BBU) are over 10 kg and 5 kg, respectively, where BBU consists of DU and CU. The average power consumption of RU and CU can reach 1.175 kWh and 0.325 kWh under a full workload, which can even reach 3.8 kWh with the three-sector setting.

\subsection{Uncrewed Aerial Vehicle}
\subsubsection{Untethered UAV} Taking DJI as an example, the flagship U-UAV platform Matrice 350 RTK achieves a loading capability of 2.7 kg with 31 minutes of flight time, which cannot carry the macro BS equipment discussed above. Although DJI has customized a delivery UAV platform with a 40 kg loading capability (i.e., DJI Flycart30). The flight time is limited to 8 minutes, which fails to guarantee continuous connections to the users. Apart from that, the battery is originally designed to support the UAV flight with a typical capacity of 0.3 kWh (e.g., TB65 from DJI). Therefore the commercial untethered UAV can only be utilized to implement a single micro or pico network with RU, DU, and CU integrated as a complete payload. 

\subsubsection{Tethered UAV}
To overcome the limitations above, the tethered UAV has been regarded as a promising solution, and it has been utilized for emergency communication in post-disaster areas. Based on the power supply provided by the tethered system, the tethered UAV achieves much stronger loading capability and endurance time, which is suitable to implement a single macro network, e.g., DG-M30 tethered UAV can carry 20 kg payloads hovering at 200 m for 24 hours.  More importantly, the optical fiber provided by the tethered system enables flexible communication payload options, including RU, RU/DU, and RU/DU/CU. However, the tethered system limits the mobility of the T-UAV, which may lead to coverage degradation.

\begin{figure*}[h!]
  \centering
  \includegraphics[scale = 0.72]{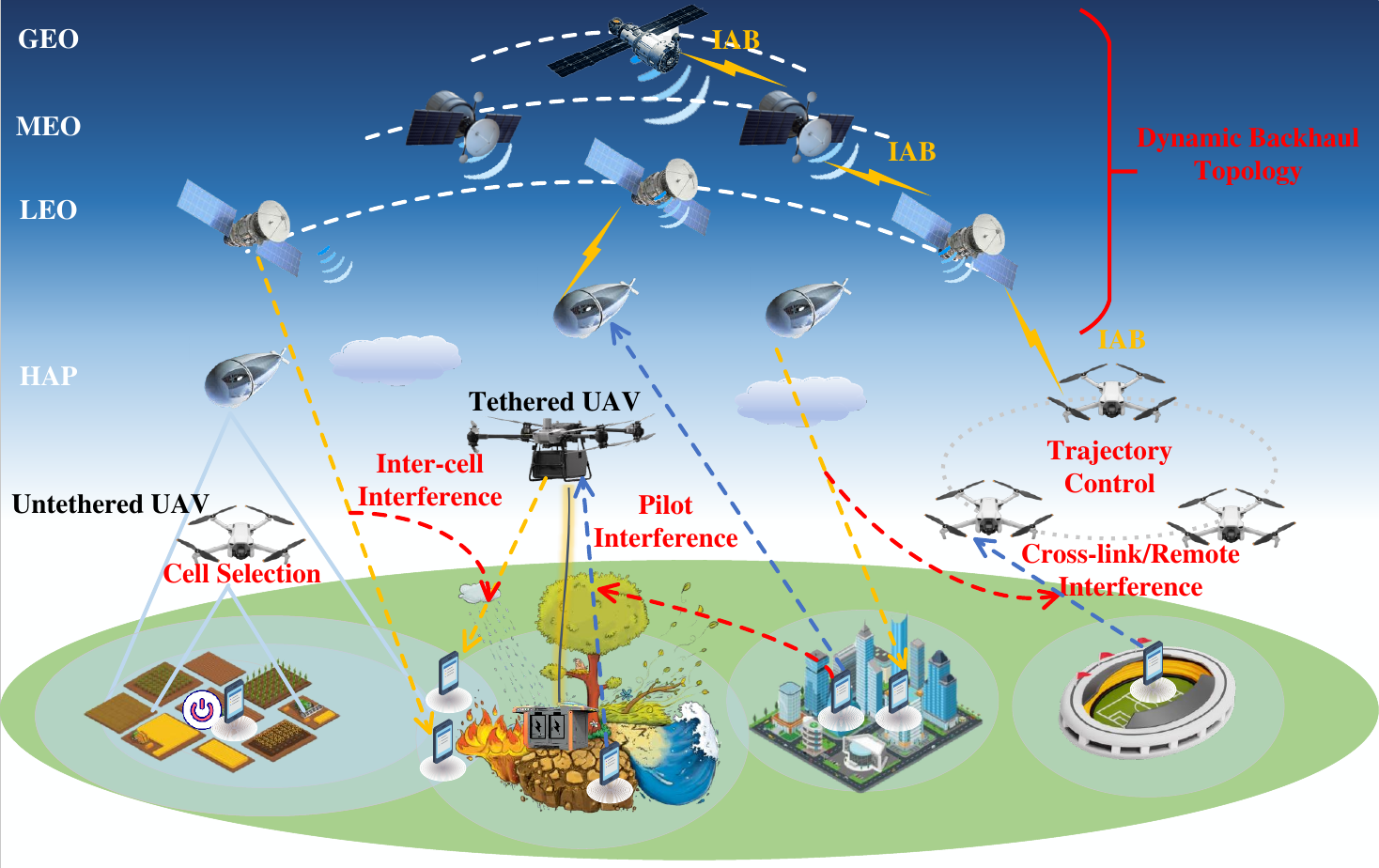}
  \caption{Individual-layer challenges in heterogeneous non-terrestrial network.}
  \label{System_model}
\end{figure*}

\subsection{High Altitude Platform}
Operated in the stratosphere e.g., 20 km, the HAP is capable of flying for several months continuously, and circles at a radius of a few kilometers or less. Taking the HAP from Stratospheric Platforms Ltd as an example, with a wingspan of 60 m, it can support a maximum of 140 kg payload and provide a 20 kW power supply. As indicated in 3GPP, the HAP can support the operation of 7 macro networks with a coverage of 100 km, where the communication payload can be RU/DU or RU/DU/CU \cite{3gppHAP}. It is noted that 3GPP supports a maximum cell range of 100 km, which means the HAP can achieve direct connectivity to the regular user terminal without modifying the protocols. More importantly, the relative stationary position of the HAP makes it a promising solution for implementing wireless backhaul or midhaul transmission based on point-to-point (P2P) LoS communication. However, it is noted that the environmental temperature of the HAP can be as low as -50 to -70 degrees Celsius. Therefore, an additional heating system should also be considered to guarantee the device's operation in a suitable temperature range.


\subsection{Satellite}

According to the operation altitude, the satellite platform can be divided into low earth orbit (LEO) satellite, medium earth orbit (MEO), and geostationary orbit (GEO) satellite. Although the GEO satellite can provide continuous connections due to its stationary position relative to the Earth, the large signal path loss and delay remain important challenges to be solved, especially for the uplink transmission. Apart from that, the GEO is exactly above the equator, which means the ground users in high-latitude areas will experience low LoS probability due to low elevation. Another common challenge for satellite communication is space weather including solar flares and geomagnetic storms, which may disrupt satellite systems and degrade the reliability of the cellular network significantly. For example, the 38 satellites of Starlink were destroyed due to the solar storm in 2022. In general, due to the advantages of less gravity and extensive sunlight, the loading capability and power supply of the satellite can guarantee the operation of 61 macro networks for years, where the Effective Isotropic Radiated Power (EIRP) reaches 59 dBW/MHz for each cell \cite{3gppNTN}. 


\section{Challenges and Coordinated Solutions in Heterogeneous Non-Terrestrial Networks} \label{section:challenges_solutions}
According to the characteristics analysis of various NTN platforms, in this section, we further present communication challenges and potential coordinated solutions of individual layers in heterogeneous NTNs as shown in Fig. \ref{System_model}, which are summarized in Table.~\ref{challenges_solutions}.

\begin{table*}[!htb]
    \caption{Challenges and Coordinated Solutions in Heterogeneous Non-terrestrial Networks}
    \begin{center}
        \begin{tabular}{|c|c|c|c|c|}
            \hline
            \textbf{Individual Layer}&\textbf{Function}  &\textbf{Heterogeneous Challenges}&\textbf{Coordinated Solutions}&\textbf{Delay-tolerant}\\
            \hline			
            \multirow{5}{*}{\makecell[l]{Physical Layer}} &Initial Access& Cell Selection &Cell Shaping&\checkmark\\
            \cline{2-5}
             &Channel Estimation&Pilot Interference& Pilot Resource Allocation&\checkmark\\
            \cline{2-5}
             &Precoding&Inter-cell Interference&Distributed MIMO&\texttimes\\
            \cline{2-5}
             &\multirow{2}{*}{Power Allocation}&Cross-Link Interference&\multirow{2}{*}{UL/DL Power Allocation}&\multirow{2}{*}{\texttimes}\\
            \cline{3-3}
            &&Remote Interference&&\\
            \hline
            \multirow{2}{*}{MAC Layer}&\multirow{2}{*}{Scheduling}&\multirow{2}{*}{Inter-cell Interference}&\multirow{2}{*}{ Coordinated Scheduling}&\multirow{2}{*}{\texttimes}\\
            &&&&\\
            \hline
            Network Layer&Routing&Dynamic Backhaul Topology&Multi-hop IAB&\checkmark\\
            \hline
            Application Layer&Trajectory Control& Distinct Control Period&Multi-timescale Control&\checkmark\\
            \hline
        \end{tabular}
        \label{challenges_solutions}
    \end{center}
\end{table*}

\subsection{Physical Layer}
\textbf{Initial Access}:
The existing initial access procedures are based on the downlink reference signal received power (RSRP) of the synchronization signal block (SSB), where the UE connects to the cell with the highest RSRP \cite{Bernab2024}. In traditional terrestrial networks, most of the cellular cells are homogeneous, where the downlink RSRP can be regarded as an effective indicator of transmission performance for cell selection. However, in the heterogeneous NTNs, cells in each tier have distinct communication capabilities, including the number of antennas, bandwidth, traffic loading, etc. Even for the NTNs in the same tier, the BSs may still have different characteristics, e.g., the tethered UAV and untethered UAV, and the visibility time length of different satellites. 

More importantly, the uplink transmission performance should also be considered during cell selection, where the UE may experience poor uplink performance to certain cells due to limited transmission power. To solve the problem, a hierarchical coordination framework should be developed for cell selection, where the NTNs perform coordinated intelligent cell shaping using optimal SSB beams, and then each UE selects the cell based on its local characteristics including transmission power. One typical example is the intelligent cell shaping developed for terrestrial networks by Ericsson, where multiple BSs coordinate to optimize the cell coverage area based on reinforcement learning. It is noted that the cell selection procedure is delay-tolerant at the Telecom operator's level. 

\textbf{Channel Estimation}:
Uplink channel estimation has been regarded as an important step in guiding downlink user data transmission, including scheduling in the MAC layer and precoding in the physical layer, where each UE is required to transmit the uplink pilot signal to the NTNs, i.e., sounding reference signal (SRS). If the UE is associated with a higher-tier NTN, higher transmission power is usually needed to compensate for the larger path loss. Therefore, in the vertical heterogeneous NTNs, UEs associated with lower-tier NTNs will experience serious pilot interference from the UEs associated with the higher-tier NTNs. Considering the much larger beam footprint in the higher-tier NTNs and limited beamforming capability at the UE, pilot interference will be more serious than that in traditional terrestrial networks.

To solve this challenge, a coordinated pilot resource allocation scheme should be developed to mitigate pilot interference and enhance uplink channel estimation accuracy based on the distribution of terrestrial UEs in each cell. It is noted that the periodicity of pilot transmission can be configured to be one of [2,5,10,20,40,80,160,320] ms, which is generally delay-tolerant in most cases.

\textbf{Precoding}:
Distributed MIMO (DMIMO) technique has been regarded as a promising solution to reduce inter-cell interference based on coherent joint transmission (CJT), which has been proven to benefit terrestrial cell-edge users significantly \cite{Haliloglu2023}. In the CJT scheme, a central server is required to obtain the information about the complete wireless channel (i.e., the wireless channel between one UE and multiple cells) and then perform the precoding algorithm to calculate the precoding matrix, such as weighted minimum mean-square error (WMMSE). It is noted that the precoding procedure is delay-sensitive and executed in every slot, e.g., 0.5 ms with 30 kHz subcarrier spacing (SCS).

However, the performance improvement of cell-edge users usually comes with the performance degradation of coordinated cells in the CJT scheme, which becomes more complicated and challenging to quantify in heterogeneous NTNs. Therefore, evaluating the overall gain of performing DMIMO and selecting the optimal coordinated cells remains an important challenge. Apart from that, the heterogeneous NTNs adopt distinct single-antenna power constraints. Hence, the challenge of guaranteeing a coherent phase with high power utilization efficiency also needs to be solved.

\textbf{Power Allocation}:
Power allocation optimization has been regarded as an effective solution to maximize the channel capacity both in single-cell and multiple-cell scenarios. In the heterogeneous NTNs, cross-link interference (CLI) and remote interference (RI) emerge as two important challenges due to the LoS-dominated aerial channel, where CLI happens between adjacent cells with different time division duplex (TDD) frame structures, and RI happens between very distant cells even with synchronized TDD frame structures. It is noted that the upper-tier NTNs usually adopt a much higher downlink transmission power than the lower-tier NTNs, which will cause strong CLI and RI to the lower-tier NTNs. To solve the CLI and RI in heterogeneous NTNs, especially for the direction from downlink (DL) to uplink (UL), a coordinated UL-DL power allocation scheme should be developed to enhance the UL signal strength and mitigate DL interference. Similar to precoding, the power allocation procedure is delay-sensitive and needs to be executed in each time slot. 

\subsection{MAC Layer}

\textbf{Scheduling}:
Different from the DMIMO technique to reduce inter-cell interference in the physical layer, the scheduling optimization has the potential to enhance the performance of cell-edge users by allocating the same resource block to UEs with near-orthogonal channel characteristics. In the heterogeneous NTNs, the multi-tier NTNs will lead to high coverage overlap, where a highly efficient coordinated scheduling algorithm is needed to mitigate the more sophisticated inter-cell interference. It is noted that the coordinated scheduling algorithm is also important for the CJT scheme in DMIMO, where the participating cells must transmit the user data on the same resource block. Otherwise, the coordinated cells will sacrifice the power consumption without improving the cell-edge UEs in the primary cell. One possible solution regarding this issue is to pre-schedule the UEs in the primary cell and share the information with coordinated cells for precoding calculation in the next slot.

\subsection{Network Layer}
\textbf{Routing}:
Since diverse mobility patterns of satellites, HAPs, and UAVs lead to dynamic backhaul topologies, it is difficult to maintain stable point-to-point LoS backhaul transmission between the NTN platforms and the core network on the ground. Therefore, the integrated access and backhaul (IAB) technology has been regarded as a promising solution for more robust backhaul transmission \cite{Lin2024}, which has been further extended to wireless access backhaul (WAB) at 3GPP Release 19. More importantly, the Backhaul Adaptation Protocol (BAP) layer is introduced in DU for routing, which makes it possible to achieve multi-hop backhaul transmission. However, the dynamic topology and heterogeneous communication capabilities make it challenging to find the optimal routing path. Therefore, the coordinated routing algorithm should be developed by considering both the local cell's status and the neighboring cells' status. Another challenge in the backhaul routing via IAB is the channel capacity of the aerial communication. Unlike traditional access links, the IAB link transmits cell-level traffic between two NTNs via an access link, which has a much higher capacity requirement. However, the propagation path is usually dominated by LoS in aerial links, which limits the application of spatial multiplexing to enhance channel capacity. One promising solution is the multi-connectivity technology to enable distributed intelligent backhaul traffic allocation on multiple routing paths. The routing table optimization is classified as a delay-tolerant function, which aims to keep the traffic relatively stable over a period of time and collect enough observations before making a new decision.

\subsection{Application Layer}
\textbf{Trajectory Control}:
Different from the terrestrial network, heterogeneous NTNs are required to be controlled to optimize the trajectories for collaboratively providing optimal coverage in large-scale areas. Although the satellites have fixed orbits based on their altitude, the satellites still can change their trajectory with a longer response time at a higher cost, e.g., in emergency communications. Therefore, due to the distinct mobility characteristics of heterogeneous NTNs, the multiple-timescale trajectory optimization algorithm should be developed to enable coordination among NTNs for optimal coverage. More importantly, the received quality of control data over wireless backhaul transmission significantly impacts the trajectory update (e.g., latency of control signal), and then degrades the performance of access links to the terrestrial UEs. Therefore, it remains an important challenge to investigate the closely coupled control signal transmission and user data transmission.

\section{MADRL-empowered Coordinated Solutions}\label{section:distributed_learning}
The MADRL algorithm has been widely considered as a promising technique to enable efficient distributed coordination in heterogeneous NTNs. Among the typical architectures, the centralized training and execution (CTE) framework suffers from high overhead and propagation delays, while the decentralized training and execution (DTE) framework struggles with non-stationarity and poor coordination. In comparison, the centralized training with decentralized execution (CTDE) framework provides a better balance between scalability and coordination \cite{Chen2021}, which is adopted as the underlying framework for the MADRL algorithm in our following discussion. Based on the delay requirements in the Table. \ref{challenges_solutions}, the delay-tolerant and delay-sensitive MADRL-empowered distributed coordination solutions are introduced as follows, respectively:

\subsection{Delay-tolerant Distributed Coordination}
According to the O-RAN structure, the near real-time (near-RT) RAN intelligent controller (RIC) deployed in the CU can typically support the RAN control with latency from 10 ms to 1 s, and the non-RT RIC deployed in the core network can support the RAN control with latency over 1 s. Therefore, for delay-tolerant distributed coordination solutions, the centralized server during training can be deployed in CU either onboard or on the ground, or even in the core network. 
\begin{itemize}
    \item \textbf{Cell Shaping}: To cooperatively provide optimal coverage in initial access, the MADRL algorithm should be developed to select the optimal SSB codebook for each cell, considering the UEs' distribution and the capabilities of neighboring cells. One potential solution is to integrate the graph neural network (GNN), which models heterogeneous NTNs as nodes with various features and helps to deal with large-scale and high-dimensional NTN scenarios, thereby facilitating the MADRL model training. Apart from that, the UEs associated with heterogeneous NTNs have distinct transmission power capabilities, which limit their connection to upper-tier NTNs such as satellite. Therefore, a lightweight model should be deployed at the UE to make the target NTN platform association decision based on local characteristics, including transmission power and battery. 
    \item \textbf{Pilot Resource Allocation}: In heterogeneous NTNs, complicated and overlapping coverage is a key cause of pilot interference. To mitigate this, the UE-level coordinated pilot resource allocation problem should be solved by the MADRL, where multiple cells learn to cooperatively minimize the channel correlation of UEs sharing the same pilot resource. However, UE-level allocation may increase the observation and action space significantly in MADRL, especially for higher-tier NTNs. Therefore, the hierarchical MADRL framework can be utilized to divide the coverage into several low-correlated smaller areas first, and then focus on the pilot resource allocation within each smaller area. To reduce the communication overhead during centralized training, channel charting can be a potential solution by extracting the low-dimensional features from the statistical channels \cite{Ferrand2021}.
     \item \textbf{Multi-hop IAB}: Although the training of the routing optimization algorithm is delay-tolerant, the user data transmission may have stringent latency requirements. Therefore, multi-hop transmission of duplicate user data leads to increased transmission latency and channel capacity demands. To solve the problem, the hierarchical goal-oriented semantic communication can be integrated with the MADRL to reduce transmission overhead by prioritizing task-relevant information in each hop \cite{zhou2024} in this heterogeneous network. Additionally, the graph encoders, such as GNN, can also
     be incorporated into each agent’s observation, enhancing the adaptability of MADRL policies to topological changes.
     \item \textbf{Multi-timescale Control}: In the multi-timescale trajectory optimization problem, the NTNs in different tiers experience different timescales of observation, action, and reward in MADRL due to varying altitude, propagation or processing delays, and control periods. Hence, a potential solution is to decouple the formulated hierarchical MADRL problem based on the timescales, and then optimize the subproblems iteratively \cite{Liu2024}. Moreover, it is noted that goal-oriented semantic communication can also be utilized to quantify the importance of control signal transmission, which is closely coupled with the performance of the access links.
\end{itemize}

\subsection{Delay-sensitive Distributed Coordination}
It is noted that the cellular network has stringent numerology and frame structure, where most of the physical layer and MAC layer procedures need to be completed within every slot length. To satisfy the stringent latency requirement of delay-sensitive coordination solutions, the central server during training should be deployed in the DU, and a single DU should connect to multiple RUs for coordination. In this case, more functions in the physical layer and MAC layer are required to be moved to the RU.
\begin{itemize}
    \item \textbf{Distributed MIMO}: Since the heterogeneous NTNs are equipped with a distinct number of antennas, the observation space (e.g., dimensions of the estimated channel) and action (e.g., dimensions of the precoding matrix) space of different agents vary from each other, the distributed MADRL algorithm with heterogeneous agents, including MADDPG, should be investigated. Considering the channel sharing overhead, to fully enable DMIMO precoding without channel sharing, one possible solution is to perform channel charting for cell-edge UEs via multiple cells in advance. Then, each NTN reconstructs the complete channel for cell-edge UEs based on the local estimated channel and channel charting map.
    \item \textbf{UL/DL Power Allocation}: In this case, each NTN is required to deploy two heterogeneous agents to optimize the UL and DL power allocation, respectively. Specifically, the first agent detects the occurrence of CLI or RI based on the received signal and increases the uplink transmission power correspondingly. Then, the second agent optimizes the downlink transmission power based on the positions of neighboring cells to mitigate causing CLI or RI. It is noted that two agents of a single NTN work sequentially in the TDD frame structure, and same type of agents in different heterogeneous NTNs may work asynchronously due to dynamic TDD. 
     \item \textbf{Coordinated Scheduling}: To reduce the inter-cell interference in the MAC layer, each agent should learn to cooperatively schedule the UEs with minimal channel correlation on the same resource block. In this case, the reward in MADRL can be designed as the sum capacity in the scenario. However, to further support the CJT scheme in DMIMO, the MADRL algorithm will incorporate the distributed scheduling consensus constraint, where the coordinated cells should schedule the cell-edge UEs on the same resource block for precoding.
\end{itemize}

\section{Case Study}\label{section:case_study}
In this section, to validate the effectiveness of the MADRL-empowered distributed coordination, we formulate a joint user scheduling and trajectory control optimization problem to maximize the average downlink throughput in a heterogeneous IAB-assisted NTN emergency communication scenario. Specifically, one T-UAV (IAB-Donor) and four U-UAVs (IAB-Node) are deployed at altitudes of 200 m and 100m, respectively. They serve the mobile ground users (G-UEs) through downlink transmission with high heterogeneity in terms of carrier frequency, bandwidth, and transmit power. During each time slot of 30 ms, we assume the incoming packets for each G-UE follow the Poisson process with a parameter of 4, and the latency for dropping packets is 10 time slots.




\begin{figure}[ht]
    \centering
        \subfigure[Overall Average downlink throughput.]
        {\includegraphics[width = 6.5cm]{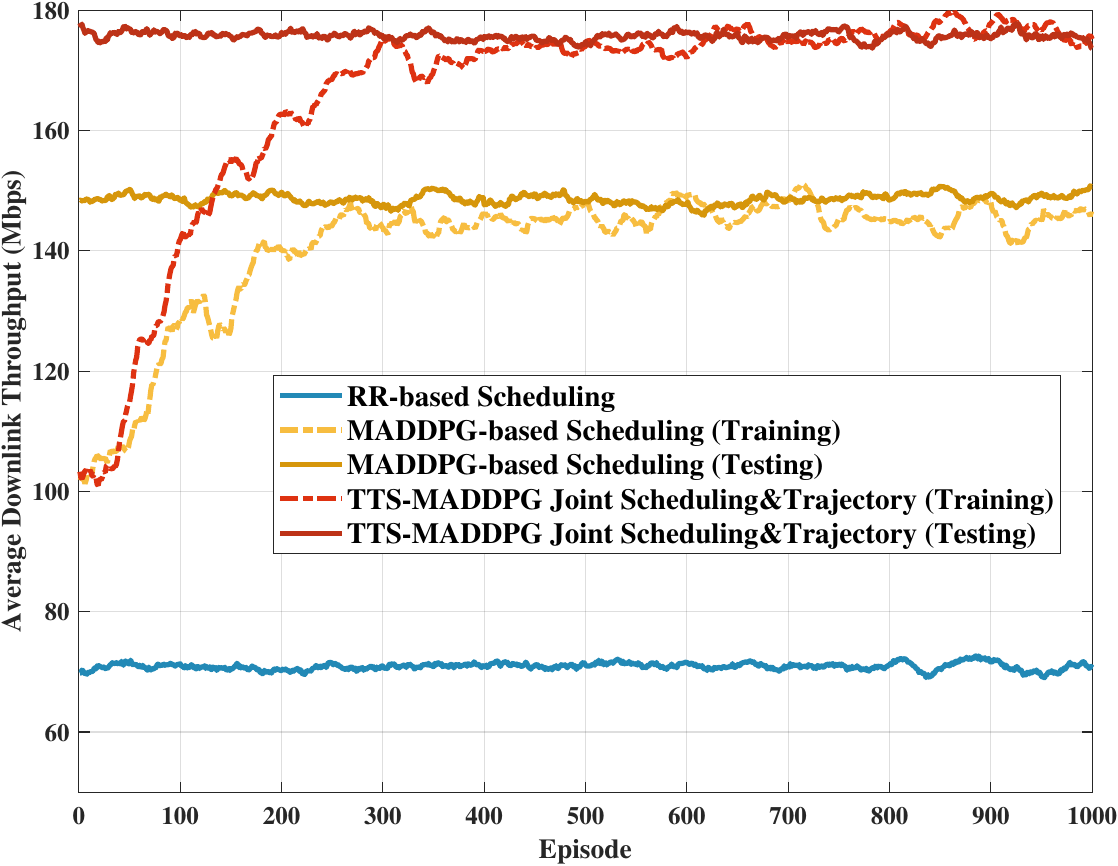}} 
        \centering
        \subfigure[Average downlink throughput of each UAV.]
        {\includegraphics[width = 6.5cm]{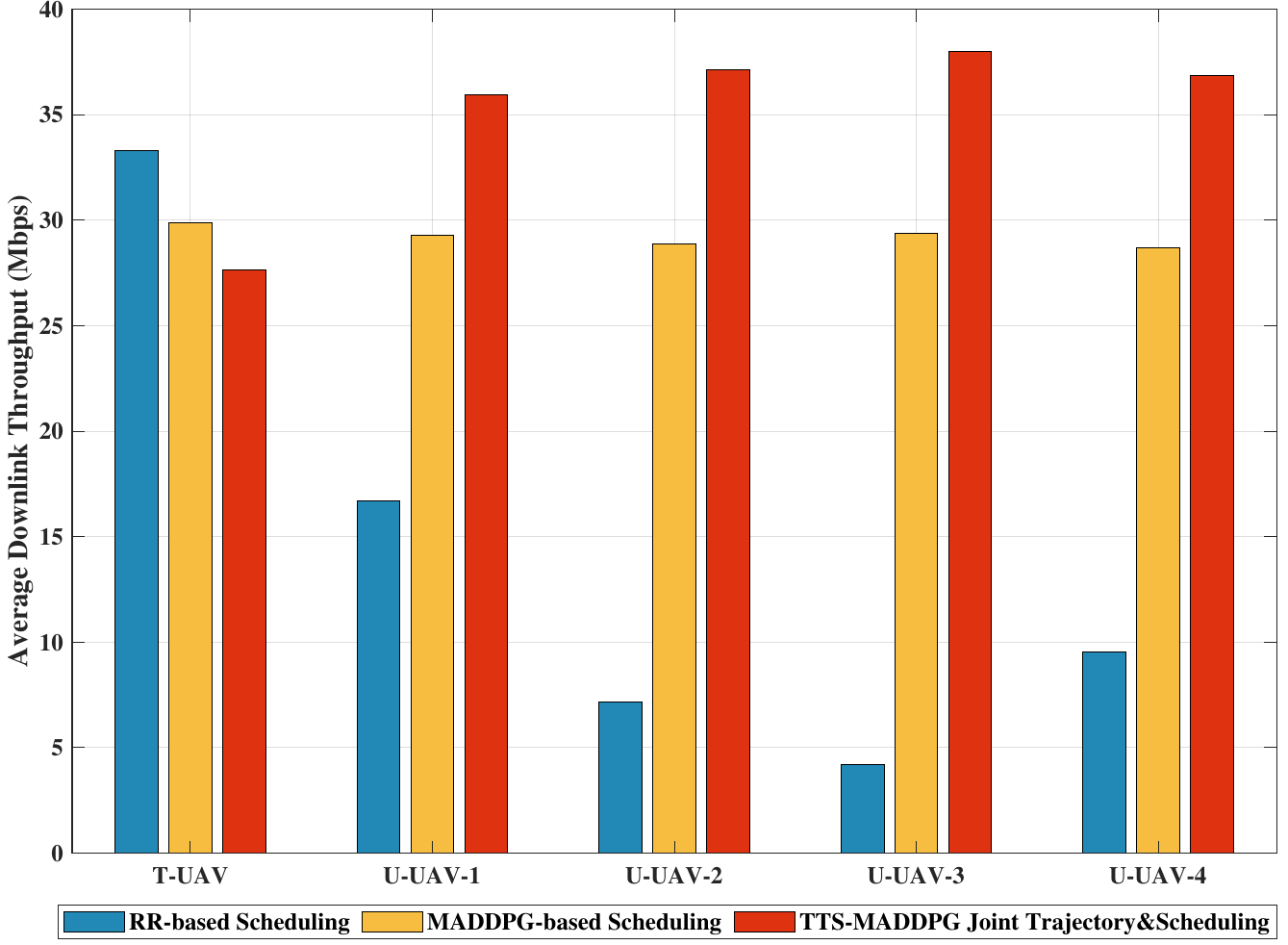}}
    \caption{Average downlink throughput performance of the proposed TTS-MADDPG algorithm}
    \label{MADDPG_performance}
\end{figure}

To solve this problem, we propose the TTS-MADDPG algorithm under the CTDE framework. Specifically, this algorithm involves two groups of agents for two tasks: the first group (all UAVs) is responsible for the short-timescale user scheduling task at each time slot, while the second group (all U-UAVs) handles the long-timescale trajectory control task at every 5 time slots due to UAV's mobility constraints. During the offline centralized training, global state information is utilized by the central critic to assist learning, while each agent’s actor is trained using only local observations.





The average downlink throughput results of our TTS-MADDPG algorithm during both training and testing phases are illustrated in Fig. \ref{MADDPG_performance}, where we also include two benchmark schemes for comparison: a round-robin (RR)-based scheduling method and a MADDPG-based scheduling approach. We conduct the simulation across 1000 episodes, with each episode consisting of 200 time slots, and the learning rates of actor and critic set to $10^{-4}$ and $10^{-3}$. In Fig. \ref{MADDPG_performance}(a), for the converged performance, the TTS-MADDPG algorithm yields approximately 150\% and 19\% throughput gains over the RR-based and the MADDPG-based scheduling benchmarks, respectively, converging at around 175 Mbps. 
During testing, the TTS-MADDPG algorithm maintains stable performance that closely matches its converged performance during training, demonstrating strong generalization capability.
In Fig. \ref{MADDPG_performance}(b), with the TTS-MADDPG algorithm, T-UAV's throughput declines slightly to 27.6 Mbps, as it learns to ensure sufficient scheduling on U-UAVs to serve the edge users and maximize the overall throughput. Moreover, each U-UAV achieves a noticeable throughput improvement of about 36 Mbps because of the optimal scheduling and real-time trajectory control.

The above results demonstrate that the proposed MADRL-based framework enables effective distributed coordination among heterogeneous UAV-based NTN platforms. While the case study focuses on the UAV-based scenario, it also exhibits rich heterogeneity in terms of physical characteristics, communication patterns, and the operation timescales, which acts as a solid foundation for extension to broader NTN architectures involving HAPs and satellites and for tackling other complicated challenges.

\section{Conclusions and Future Work}\label{section:conclusion}
In this paper, we summarize the unique characteristics of different NTN platforms and analyze their impact on the implementations. 
We then focus on investigating the communication challenges for heterogeneous NTNs in individual layers, along with their potential coordinated solutions. 
We further investigate the distributed MADRL algorithms tailored for each potential coordinated solution in heterogeneous NTNs, and analyze the CTDE deployment options according to their latency requirement.
Importantly, we utilize the TTS-MADDPG algorithm to solve the joint trajectory design and user scheduling optimization problem in a heterogeneous NTN composed of both T-UAV and U-UAV, which validates the effectiveness of the distributed coordination solutions. 

Moreover, in the future extended NTN scenario with a vast number of agents and multiple timescales, it is challenging to achieve effective distributed coordination, leading to slow and unstable convergence in the MADRL training. To address these issues, the MADRL can incorporate goal-oriented semantic communication and the graph encoders to reduce transmission overhead, computational load, and enhance the convergence. Moreover, it can also be extended to asynchronous MADRL based on the DTE framework, or further integrated with the grouped training and decentralized execution (GTDE) framework to improve scalability. 







\vspace{-2\baselineskip}
\begin{IEEEbiographynophoto}{Jikang Deng} 
is currently a Ph.D. student at King Abdullah University of Science and Technology (KAUST), Jeddah, Saudi Arabia. He received the B.Eng. degree in network engineering from the University of Electronic Science and Technology of China (UESTC), Chengdu, China, in 2023. His research interests include wireless communication, non-terrestrial networks, and AI for communication.
\end{IEEEbiographynophoto}
\vspace{-2\baselineskip}
\begin{IEEEbiographynophoto}{Hui Zhou} is currently an Assistant Professor at Coventry University. Before joining Coventry University, he held research positions at King Abdullah University of Science and Technology, Saudi Arabia, and Huawei, China. His research interests include non-terrestrial networks, integrated sensing and communication, unlicensed spectrum, machine learning, and semantic communication.
\end{IEEEbiographynophoto}
\vspace{-2\baselineskip}
\begin{IEEEbiographynophoto}{Mohamed-Slim Alouini} is currently the Al-Khawarizmi Distinguished Professor of Electrical and Computer Engineering at King Abdullah University of Science and Technology (KAUST) and the holder of the UNESCO Chair on Education to Connect the Unconnected. He is a Fellow of the IEEE and OPTICA, and his research interests include the modeling, design, and performance analysis of wireless communication systems.
\end{IEEEbiographynophoto}	
	

\end{document}